# Free space laser telecommunication through fog

GUILLAUME SCHIMMEL,[1] THOMAS PRODUIT,[1] DENIS MONGIN,[1] JÉRÔME KASPARIAN,[1,2] AND JEAN-PIERRE WOLF[1,*]

[1]Group of Applied Physics, University of Geneva, Chemin de Pinchat 22, 1211 Geneva 4, Switzerland
[2]Institute for Environmental Sciences, University of Geneva, bd Carl Vogt 66, 1211 Geneva 4, Switzerland
*Corresponding author: Jean-Pierre.Wolf@unige.ch

**Atmospheric clearness is a key issue for free space optical communications (FSO). We present the first active method to achieve FSO through clouds and fog, using ultrashort high intensity laser filaments. The laser filaments opto-mechanically expel the droplets out of the beam and create a cleared channel for transmitting high bit rate telecom data at 1.55 µm. The low energy required for the process allows considering applications to Earth-satellite FSO and secure ground based optical communication, with classical or quantum protocols.**

Significant efforts are currently dedicated to establish free space optical communications (FSO) [1], both classical [2–5] and quantum [6–10], within a network of satellites [11, 12], between the Earth and satellites [5, 10] and between the Earth and drones [13]. For instance, the first free space quantum communication satellite from the Chinese Aerospace Agency demonstrated quantum key distribution over 1200 km between the satellite and the Earth [9]. Space agencies also support laser telecommunication programs [14] aiming at significantly increasing the data rates as compared to radio-frequencies (RF) and reducing the problem of the availability of RF bands.

The major limitation for optical free space telecommunication is the availability of a clear sky, i.e. free from fog and clouds. This critical issue is currently only addressed by the multiplication of networked ground stations, which is complex and expensive. No active methods, with the objective of creating a clear channel within clouds or fogs to allow data transmission, have been proposed and implemented so far. This is the aim of the present paper.

Early attempts to clear the sky from fog and clouds with high power $CO_2$ lasers took place in the 70's and 80's, mainly for increasing visibility on the battlefield. However, the very high energy required to vaporize and shatter water drops (lasers of typically 10 kW.cm$^{-2}$ continuous wave [15] and 10-1000 MW.cm$^{-2}$ pulsed [16, 17]) was prohibitive for applications with fogs extending over 100 m thickness. Moreover, the laser energy was deposited in the first meters of the optical path, according to the usual Beer-Lambert exponential decay.

The 2000's saw the emergence of femtosecond TW-class lasers (1 TW = $10^{12}$ W), hence the opportunity to reconsider laser transmission through fog with a fundamentally different approach: non-linear propagation in the atmosphere and laser filamentation [18]. Due to the high peak-power, focusing (Kerr effect) and defocusing (Kerr saturation, plasma generation) non-linearities involved in the propagation lead to the formation of laser filaments [19–23]. Laser filaments are self-sustained light structures of typically 100 µm diameter (at 800 nm) and up to hundreds of meters in length, widely extending the traditional linear diffraction limit. They bear high intensities (10–100 TW.cm$^{-2}$) and generate a low density plasma in air with typically $10^{16}$ charges per cm$^3$. A laser filament also survives interaction with water droplets [24–27]. The energy diffracted by the droplet [28] feeds the surrounding beam, which allows Kerr re-focusing and thus a fast healing of the filamentary structure. However, Mie scattering during propagation through the fog reduces the energy reservoir of the whole beam around the filament, and restricts the self-healing process after interaction with a large number of water drops.

In the present paper, we demonstrate a radically new approach, which consists in opto-mechanically displacing the fog droplets instead of directly interacting with them. When a filament is created in air, the plasma generation produces a shockwave that leads to a reduced density channel significantly larger than the filament itself, and over a time window of 0.1-1 ms [29-33]. While launched in an ensemble of droplets, this filament-induced shockwave radially expels the droplets out of the beam from the air it sweeps. In contrast to e.g., droplet evaporation, the shockwave can therefore clear the air well beyond the volume of filament itself. If the repetition rate is faster than the time required for the particles to re-enter the cleared volume, this volume forms a particle-clear channel with a mm-range diameter in a quasi-stationary regime [34]. However, the TW-class laser beam is of limited value to carry high bit-rate telecom data, as its repetition rate lies in the kilohertz range.

We show that the volume cleared from water droplets by the filament can be used to transmit telecom data through the cloud, by coupling a second laser into the channel. We also show that the data bandwidth of the channel is unlimited. The high peak power laser creates a local cloudless pathway with reduced Mie scattering for the second, modulated, telecom beam that carries information. The spatial overlap between the two laser beams is of critical importance, both for the transmission enhancement and for the reduction of thermal lensing induced by the filament. Two regimes are observed in the present proof of principle experiment. At low repetition rate, the filaments open a short time window - tens of millisecond - where the transmission of the data-carrier laser is increased. At high repetition rate (100-1000 Hz), the filaments establish a quasi steady-state regime where droplets are systematically forced out the channel, so that the telecom laser transmission shows a spectacular and quasi-continuous increase. While adaptive optics beam shaping has been used to improve performance for FSO [35], it has no influence on the atmosphere itself, nor on the beam transmission. Therefore, our experiment constitutes the first active method opening an optical link through fogs and clouds.

The demonstration experiments (see Materials and Methods in supplementary materials and Fig. S1) are performed on a laboratory cloud chamber with controlled parameters, such as water droplet size distribution, number density, humidity and temperature. In order to simulate realistic cloud conditions, the size distribution of the droplets is centered about 5 μm and the optical attenuation modified from 0 (no cloud) to 20 dB. To this end the droplet number density was set from 0 to $10^4$ cm$^{-3}$ (100 times more than in a cumulus cloud). The laser filaments are produced by 30 fs, 800 nm, 5 mJ laser pulses slightly focused in the cloud chamber with a 1 m focal-length lens to form 20 cm-long filaments. The telecom laser is superimposed to the high intensity laser in a counter-propagating geometry (see Fig. S1). In order to investigate the dynamics of the droplets expulsion by the shockwave, experiments have been first carried out with a continuous wave (cw) laser transmitter (Fig. 1). The average transmission through the fog shows a strong dependence on the repetition rate of the high-power laser (Fig. S3), which generates the filaments and opens the cloudless channel. In absence of a well-defined multiple filament arrangement like in [33], the previously mentioned channel represents a region with a reduced water droplets concentration. At low repetition rates (≤ 50 Hz), the time interval between the pulses is longer than the time required by the droplets to refill the channel so that, on average, the transmission is comparable to the case without filament. However, about 200 μs after the pulse, the transmission exhibits an increase (in this example about 15% for an initial cloud transmission of -5.6 dB, see the 10 Hz curve), before a slow decay of 10 ms. An operative time window, in which data can be received with increased transmission, is thus opened after the femtosecond pulse, with a typical half time $T_{1/2}$ = 5 ms. At repetition rates higher than 200 Hz, corresponding to pulse intervals shorter than $T_{1/2}$, the transmission of the continuous wave laser exhibits a dramatic increase: Even for clouds with initial attenuation larger than 10 dB, the transmission reaches 95% beyond 500 Hz. At this time interval between pulses, the channel reaches a quasi-steady state regime where nearly no droplet is present in the channel anymore. This quasi-stationary regime is achieved by a cumulative effect of the pulses, i.e. successive ejections of the droplets with a time interval smaller than the time $T_{1/2}$ needed to refill the channel. A repetition rate above 200 Hz is therefore optimal for opening an information transmission channel. These results constitute the first experimental demonstration of the opto-mechanical expulsion model proposed previously [34].

Interestingly, right after the filament shockwave, a fast and short transmission drop of about 10 to 15% is observed. These initial transient losses induced by the filament are discussed in detail in the supplementary material. Briefly, they result from the air refractive index gradient, associated with the heat deposited by the filament [29-33]: the gas-depleted region exhibits a lower refractive index at the center, which induces a defocusing effect here on the telecom laser.

The efficient expulsion of the fog droplets results in a drastic reduction of Mie scattering, as illustrated in Fig 2 and the associated movies (supplementary material). The four pictures show the dramatic reduction of linear Mie scattering on the side of the cloud when the clearing filaments (at 1 kHz) are absent (A) or present (B), as well as the related increase in transmission (C and D).

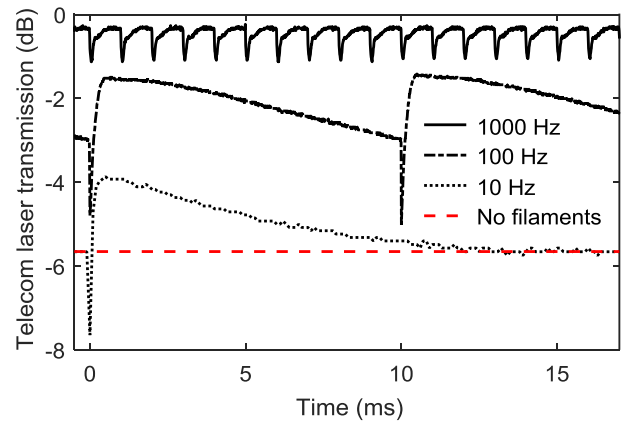

**Fig. 1.** Time evolution of the optical transmission of the telecom laser through fog (producing an attenuation of 5.6 dB) for three different repetition rates of the high intensity laser and without filament. Notice the decrease of the optical attenuation (i.e. increase of average transmission) with the repetition rate and the thermal defocusing effect that occurs dring the first hundreds of microseconds after each pulse followed by an operative time window.

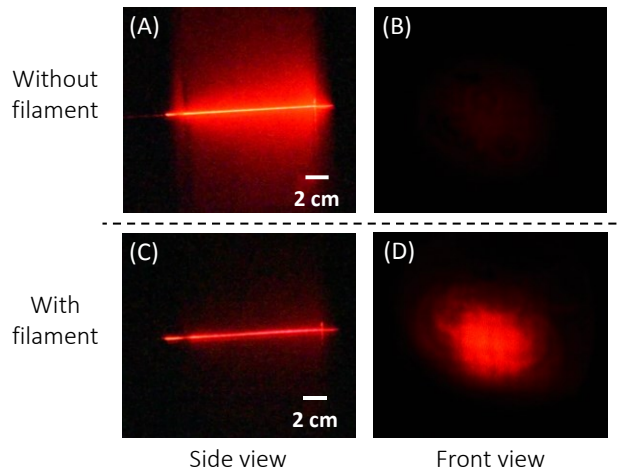

**Fig. 2.** Frames from the Supplementary videos displaying the data-carrier laser transmission (visible 633 nm laser in this case) through fog with and without filament (1 kHz). (A,C): side view (see movie S1); (B,D): front view (see movie S2). Front and side views with or without filament are respectively captured with the same video camera settings.

The demonstration of free space optical telecommunication through fog and clouds was performed with a HF modulated telecom transmitter, at the standard wavelength for earth-satellite transmission, i.e. 1.55 μm. While in principle, no limitation on the channel bandwidth is expected as the channel stays open in a quasi-steady state, the demonstration focuses on modulations up to 1 GHz (see Fig. S4). To take advantage of the channel cleared by the filament shockwave, the coupling of the auxiliary laser into the channel is a key parameter. The beam waist of the telecom laser must be sufficiently small along the whole cleared channel to benefit from the droplets expulsion. It implies to focus the beam with a relatively large f-number $f_\#$ (i.e. weak-focusing). Experimentally, we found the best focusing parameters to be $f_\# = 150$ with 1 m focal length, representing a good trade-off between maintaining a small beam radius (< 300 μm) along the channel to avoid scattering and keeping a strong enough focusing ($f_\# < 200$) to limit the defocusing by the thermal gradient created by the filament. With $f_\# = 150$, the beam radius was lower than 300 μm (at $e^{-2}$) all along the 20 cm cleared channel, which provides a lower-bound estimation of the cleared channel radius regarding the improvement of the beam transmission.

Figure 3 demonstrates the ability of transmitting 0.2 GHz data rates through a dense cloud with -12 dB initial transmission, by the filament clearing method. While, without filaments, virtually no information is transmitted, 80% transmission with full modulation depth is achieved through the cleared channel. As expected from the experiment with the cw laser, no bandwidth limitation is observed, and similar results are obtained from 1 MHz to 1 GHz, which are the limits of our data transmitters/receivers (see Materials and Methods in supplementary material).

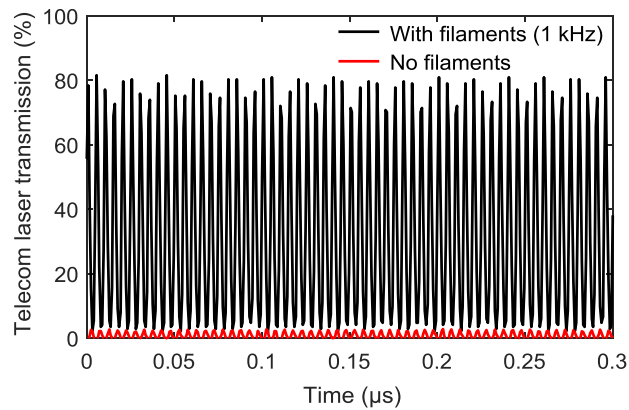

**Fig. 3.** Transmission of the 0.2 GHz amplitude modulated telecom laser through fog (producing an attenuation of 12 dB) as a function of time, with and without laser filament (1 kHz) through fog.

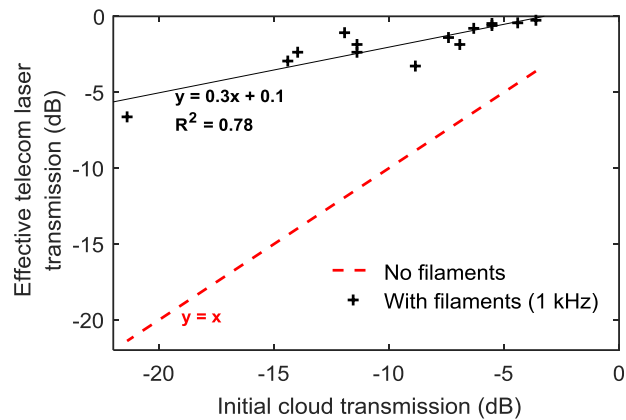

**Fig. 4.** Increase of the cloud optical transmission for the data carrier telecom laser, allowed by the laser filaments.

The spectacular transmission enhancement allowed by the laser filaments can be translated in a reduction of the cloud attenuation in dB, as shown in Fig. 4. To this end, the number density of the cloud droplets was modified in order to obtain clouds with initial transmission ranging from 0 dB to -20 dB. The measurements show that, interestingly, the gain produced by the filament induced shockwave is not constant in dB, but increases with the initial cloud thickness. This higher cleaning efficiency at higher droplet densities can be explained by the fact that the shock wave displaces a defined air volume, regardless of the concentration of droplets contained in this volume. Therefore, if the volume contains a larger number of droplets, the shock wave induces a larger transmission increase. In other words, the energy from the laser is mainly used to heat the air molecules that produce the shockwave, and the kinetic energy from the displaced droplets is negligible in comparison, for any reasonable conditions (less than $10^4$ droplets/cm$^3$).

Note that clearing the air from droplets over long distances does not require a single continuous laser filament. Rather, successive filaments or a filament bunch are sufficient to create a shockwave over the whole filamentation length and therefore to open a continuous droplet-clear channel. As multifilamentation over hundreds of meters has been widely demonstrated [36, 37], generating the cleared channel would not be the limiting factor for long-distance applications. In the transverse direction, droplet are expelled at a sufficient speed to quickly leave the filamenting region, avoiding any local accumulation in regions between filaments. Specific filament arrangements, as proposed in [33], might influence the efficiency of the channel clearing by adding waveguiding, although such optimization is well beyond the scope of the present work.

Application to real-scale fog and clouds extending over long distance also implies maintaining a small enough beam diameter for the telecom laser along the cleared channel. The channel diameter measured in the present experiment is in the range of one millimeter (greater than 600 µm), but could be increased if more laser energy was available and deposited in the plasma. An attractive option consists in considering mid-IR filaments instead NIR. Indeed, mid-IR filament exhibit a much larger diameter (about 4 cm at 10 µm [42]) and are expected to span over considerable distances [38-43].

Another key aspect for real scale application is the energy loss due to the creation of the plasma induced shock wave. The deposited energy required for such a shockwave formation associated with air ionization is of the order of 0.1 to 0.4 mJ/m at 800 nm [43], for pulse powers of few times the critical power for self-focusing. On kilometric distances, the required energy would therefore range from 100 mJ to 1 J at 1 kHz. Such lasers are currently in development, thanks to the remarkable progress in the thin disk laser technology [44, 45].

The presented experiment put forward the existence of an operative time window for communication even when using a low repetition rate high-power laser. However, the best results were obtained at kilohertz repetition rate that creates a quasi steady state regime with a 30 fold increased optical transmission through fog, and thus a stationary high speed communication channel. Our experiment has been carried on a laboratory-scale fog length, with a typical water droplet density more than 100 times higher than real fog or cloud, hence an optical density comparable to realistic atmospheric conditions (for clouds, between 10 to 100 dB/km). Nevertheless, some significant issues need to be addressed to extrapolate our results to atmospheric ranges, such as the beam divergence of the data-transmitter over hundreds of meter cloud thickness or atmospheric turbulence [46].

**Funding.** ERC advanced grant "Filatmo"; Swiss National Science Foundation SNSF (200021_178926). TP acknowledges the European Union's Horizon 2020 research and innovation programme (737033-LLR).

**Acknowledgment**. We gratefully acknowledge support from Michel Moret.

See Supplement 1 for supporting content.